%% file: JournalSCFSCP.tex
\documentclass[10pt]{article}

\usepackage[utf8]{inputenc}

\usepackage[T1]{fontenc}
\usepackage[normalem]{ulem}
\usepackage[english,french]{babel}

\usepackage{verbatim}
\usepackage{graphicx}
\usepackage{latexsym}

\usepackage{graphics}
\usepackage{epsfig}
\usepackage{amsmath}
\usepackage{rotating}
\usepackage{url}
\usepackage{fancyhdr}

\usepackage{tabularx}
\usepackage{array}
\usepackage{enumitem}
\usepackage{amsthm}
\usepackage{amsmath}
\usepackage{amssymb}
\usepackage{makeidx}
\usepackage{multirow}

\usepackage{multicol}

\usepackage{float}
\usepackage[margin=1.35in]{geometry}
\usepackage{tikz}
\usepackage{pstricks}
\usepackage{times}
\usetikzlibrary{trees}

\date{}

\theoremstyle{definition}
\newtheorem{thm}{Theorem}[section]

\newtheorem{lem}[thm]{Lemma}
\newtheorem{pre}[thm]{Proposition}

\newtheorem{defini}[thm]{Definition}
\newtheorem{exemple}[thm]{Example}

\newtheorem{preuve}{Proof}
\newtheorem*{preuve2}{Proof}

\title{\huge{Relaxed Conditions for Secrecy in a Role-Based Specification}}
\author{Jaouhar Fattahi, Mohamed Mejri and Hanane Houmani}

\begin{document}                               

\makeatletter
  \begin{titlepage}
  \centering

   \vspace{1cm}
      {\large\textbf{	\@date}}\\
    \vspace{1cm}
       {\large \textbf{\@title}} \\
    \vspace{1cm}

\begin{tabular}{ccc}
         {\large \textbf{Jaouhar Fattahi}$^{a}$}  &    {\large \textbf{Mohamed Mejri}$^{a}$}  &        {\large \textbf{Hanane Houmani}$^{b}$} \\

       {\textit{jaouhar.fattahi.1@ulaval.ca}} &   {\textit{mohamed.mejri@ift.ulaval.ca}}  & {\textit{hanane.houmani@ift.ulaval.ca}}\\ 

\end{tabular}

    \vspace{1em}
$^{a}$ {\textit{LSI Group, Laval University, Quebec, Canada}}\\
$^{b}$ {\textit{University Hassan II, Morocco}}

    \vspace{1em}
        {\large \textbf{}} \\ 

\selectlanguage{english}
\begin{abstract}

 \noindent  

In this paper, we look at the property of secrecy through the growth of the protocol. Intuitively, an increasing protocol preserves the secret. For that,  we need functions to estimate the security of messages. Here, we give relaxed conditions on the functions and on the protocol and we prove that an increasing protocol is correct when analyzed with functions that meet these conditions.\\
$ $

\end{abstract}
  \end{titlepage}

\thispagestyle{empty}

\selectlanguage{english}
	\input{motivation}

          \input{notation}

\input{Confpreuve}

\input{ConclusionLeger}

\selectlanguage{english}
\bibliographystyle{unsrt}
\bibliography{Ma_these}

\normalsize
 \include{Appendix}

\end{document}

%% file: motivation.tex
\section{Introduction}
Checking  security in cryptographic protocols is a hard task~\cite{Rusinowitch1,Lowe2,BAN4,Durgin1,Lundh1,Meadows2011}. The problem is in general undecidable~\cite{Durgin1,Lundh1,Combinaison-Eq-journal,CortierSurvey1,Ref42}. However, a number of semi-decidable verification methods have emerged in the last decades working on several axes~\cite{BAN3,ModelCheking3,games1,games3,AbadiTyping,Debbabi33,Blanchet1,Gordon_authenticityby}. Others are decidable but under strict conditions~\cite{Ref93,Ref94,Shmatikov04decidableanalysis}. In this paper, we look at the property of secrecy in a protocol by observing its monotony. By intuition, an increasing protocol protects its secret inputs. This means, if the security of every atomic message does not decrease during its life cycle in the protocol (between all receiving and sending steps), the secret is never uncovered. For that, we have to define safe metrics to reasonably estimate the security of any atomic message. This way of approaching secrecy in protocols has been adopted in  few prior works. For instance, Steve Schneider in~\cite{Schneider4}  proposed the rank-functions as safe metrics to verify protocols in CSP~\cite{Schneider96,SchneiderD04}. These functions successfully managed to analyze several classical protocols like the Needham-Schroeder protocol. However, such verification dictates the protocol implementation in CSP. Besides, building rank-functions for every protocol is not an effortless task and their existence is not guaranteed~\cite{Heather}. In~\cite{AbadiTyping} Abadi, by using Spi-Calculus~\cite{SPI1,SPI2}, stipulates that: "If a protocol typechecks, then it keeps its secrets". For that, he forces the exchanged messages to have formally the following types: \{secret, public, any, confounder\} in order to easily estimate the security of every message from its type. 
Although this approach is simple and elegant, it cannot verify protocols that had been implemented with no respect to this logic. Alike, Houmani et  al.~\cite{Houmani1,Houmani3,Houmani8,Houmani5} defined universal functions called interpretation functions that were able to verify  protocols statically. They operate on an abstraction of the protocol called generalized roles in a role-based specification~\cite{Debbabi11,Debbabi22,Debbabi33}. An interpretation function must meet few conditions to be considered safe for analysis. Obviously, less we have restrictions on functions, easier we can build instances of them and more we have possibilities to prove protocols secure since one function might fail to prove the security of a protocol but another may succeed. In this regard, we notice that the conditions on the Houmanis' interpretation functions were so restrictive that very few functions were proposed in practice and proved safe. In this respect,  
we believe that the condition of full-invariance by substitution, which is a  property key that allows any decision made on messages of the generalized roles (messages with variables) to be exported to valid traces (closed messages), is the most restrictive one.
Since the aim of our approach is to build several functions, we think that if we release  functions from this condition  we can reach this goal.

%% file: notation.tex
\section{Notations}
Here, we give some conventions that we use in this paper. 
\begin{itemize}
\item[+] We denote by  ${\cal{C}}=\langle{\cal{M}},\xi,\models,{\cal{K}},{\cal{L}}^\sqsupseteq,\ulcorner.\urcorner\rangle$ the context of verification including the parameters that affect a protocol analysis:
\begin{itemize}
\item[$\bullet$] ${\cal{M}}$ : is a set of messages built from the algebraic signature $\langle\cal{N}$,$\Sigma\rangle$ where ${\cal{N}}$ is a set of atomic names (nonces, keys, principals, etc.) and $\Sigma$ is a set of defined functions ($enc$:\!: encryption\index{Encryption}, $dec$:\!: decryption\index{Décryption}, $pair$:\!: concatenation (denoted by "." here), etc.). i.e. ${\cal{M}}=T_{\langle{\cal{N}},\Sigma\rangle}({\cal{X}})$.  We denote by $\Gamma$  the set of all substitutions from $ {\cal{X}} \rightarrow {\cal{M}}$.
We denote by $\cal{A}$ the atomic messages in ${\cal{M}},$  by ${\cal{A}}(m)$ the set of atomic messages (atoms) in $m$ and by ${\cal{I}}$ the set of agents (principals) including the intruder $I$. We denote by $k{^{-1}}$ the reverse form of a key $k$ and we assume that $({k^{-1}})^{-1}=k$.
\item[$\bullet$] $\xi$ : is the equational theory that expresses the algebraic properties of the functions defined in $\Sigma$ by equations. e.g. $dec(enc(x,y),y^{-1})=x$. 
\item[$\bullet$] $\models_{\cal{C}}$ : is the inference system of the intruder under the equational theory $\xi$. Let $M$ be a set of messages and $m$ be a message. $M$ $\models_{\cal{C}}$ $m$ expresses that the intruder can infer $m$ from $M$ using her capacity. We extend this notation to valid traces as follows: $\rho$ $\models_{\cal{C}}$ $m$ means that the intruder can  deduce $m$ from the messages in the trace $\rho$.
We assume that the intruder has the full control of the net as described in the Dolev-Yao model~\cite{DolevY1}. She may redirect, delete and modify any message. She holds the public keys of all participants, her private keys and the keys that she shares with other participants. She can encrypt or decrypt any message with the keys that she holds. Formally, the intruder has generically the following rules for building messages:\\

\begin{itemize}
\small{
  \item[] $(int):$ $\frac{\square}{M\models_{\cal{C}} m}[m\in M\cup K(I)] $\\

  \item[] $(op):$$\frac{M\models_{\cal{C}} m_1,..., M\models_{\cal{C}} m_n }{M\models_{\cal{C}} f(m_1,...,m_n)}[f\in \Sigma]$\\

  \item[] $(eq):$$\frac{M\models_{\cal{C}} m', m'=_{\cal{C}} m}{M \models_{\cal{C}} m}$, with  $(m'=_{\cal{C}} m\equiv m'=_{\xi_{({\cal{C}})}} m)$\\
}
\normalsize
\end{itemize}

\item[$\bullet$] ${\cal{K}}$ : is a function from ${\cal{I}}$ to ${\cal{M}}$. It returns for an agent a set of atomic messages describing her initial knowledge. We denote by $K_{{\cal{C}}}(I)$ the initial knowledge of the intruder, or just $K(I)$ where the context is clear.
\item[$\bullet$] ${\cal{L}}^\sqsupseteq$ : is the lattice of security $({\cal{L}},\sqsupseteq, \sqcup, \sqcap, \bot,\top)$ that we use to assign security levels to  messages. 
An example of a lattice is $ (2^{\cal{I}},\subseteq,\cap,\cup,\cal{I}, \emptyset)$ that we use in this paper. 
\item[$\bullet$] $\ulcorner .\urcorner$ : is a partial function that attributes a level of security (type) to a message in ${\cal{M}}$. Let $m$ be a message and $M$ be a set of messages. We write $\ulcorner M \urcorner \sqsupseteq \ulcorner m \urcorner$ when
$\exists m' \in M. \ulcorner m' \urcorner \sqsupseteq \ulcorner m \urcorner$
\end{itemize}
\item[+]
Our analysis operates in a role-based specification. A role-based specification is a set of generalized roles. A generalized  role is  a  protocol  abstraction  where  the emphasis is put on some principal and
where all the unknown messages, that could not be verified, are replaced by variables. An  exponent  $i$  (the
session  identifier)  is  added  to a fresh message to express that this component
changes values from one execution to another. A generalized role expresses how an agent sees and understands the exchanged messages. A generalized role
may be extracted from a protocol by these steps:
    \begin{enumerate}
    \item We extract the  roles from a protocol.
   \item  We replace the unknown messages by fresh variables in each role.
    \end{enumerate}
Roles  can  be extracted following these steps:

\begin{enumerate}
\item For every principal, we extract  all  the  steps  in  which  she
participates.  Then, we add  a session identifier $i$ in the steps identifiers and in fresh values. 

For example, from  the variation of the Woo-Lam protocol given in Table~\ref{Wooo}, we extract three roles, denoted by $R_A$ (for the agent  $A$), $R_B$ (for the agent $B$), and $R_S$  (for  the server  $S$).

\item We introduce an intruder  $I$  to express the fact that  received messages and sent messages are perhaps received or sent by an intruder.

\item Finally, we extract all  prefixes from these roles. A prefix must end by a sending step. 

\end{enumerate}

\begin{center}
\begin {table}[H]
\begin{center}
\begin{tabular}{|ll|}
     \hline
     &\\
                    $p$   =&$\langle 1,A\rightarrow B: A\rangle.$  \\
                    & $\langle 2,B\rightarrow A: N_b\rangle.$ \\
                    & $\langle 3,A\rightarrow B: \{ N_b.k_{ab}\}_{k_{as}}\rangle.$ \\
                    & $\langle 4,B\rightarrow S:\{A.\{N_b.k_{ab}\}_{k_{as}}\}_{k_{bs}}\rangle.$  \\
                    & $\langle 5,S\rightarrow B:\{N_b.k_{ab}\}_{k_{bs}}\rangle$.\\ &\\
              \hline  
\end{tabular}
\caption{The Woo-Lam Protocol}
\label{Wooo} 
\end{center}
\end {table}
\end{center}

            From the roles, we define the generalized roles. A generalized role  is
            an  abstraction  of  a role where unknown messages are converted to variables.
            Indeed, a message or a component of a message is replaced by a variable  when
            the receiver cannot make any verification on it and so she is not sure about its integrity or its origin. The generalized roles  give  an accurate  idea  on  the behavior and the knowledge of  participants during the protocol execution.
           The generalized roles of $A$ are:
\small{
           \[\begin{array}{l}\begin{array}{lllllll}
                    {\cal A}_G ^1 =& \langle  i.1,&  A    & \rightarrow & I(B)&  :&  A \rangle
                    \end{array}\\
                    \\
                    \begin{array}{lllllll}
                    {\cal A}_G ^2=& \langle  i.1,&  A    & \rightarrow & I(B)&  :&  A \rangle .\\
                    & \langle i.2,&  I(B) & \rightarrow & A   &  :&  {X} \rangle .\\
                    & \langle i.3,&  A    & \rightarrow & I(B)&  :&  \{{X}.k_{ab}^i\}_{k_{as}}\rangle
                    \end{array}
             \end{array}\]
}
\normalsize
            The generalized roles of $B$ are:
\small{
            \[\begin{array}{l}\begin{array}{lllllll}
                {\cal B}_G ^1=& \langle i.1,&  I(A) & \rightarrow & B   &  :&  A \rangle .\\
                        & \langle i.2,&  B    & \rightarrow & I(A)&  :&  N_b \rangle  \\
                    \end{array}\\
                    \\
                    \begin{array}{lllllll}
                {\cal B}_G ^2=& \langle i.1,&  I(A) & \rightarrow & B   &  :&  A \rangle .\\
                        & \langle i.2,&  B    & \rightarrow & I(A)&  :&  N_b \rangle .\\
                        & \langle i.3,&  I(A) & \rightarrow & B   &  :&  {Y} \rangle .\\
                        & \langle i.4,&  B    & \rightarrow & I(S)&  :& \{A.{Y} \}_{k_{bs}}\rangle  \\
                    \end{array}\\
                    \\
                    \begin{array}{lllllll}
                {\cal B}_G ^3=& \langle i.1,&  I(A) & \rightarrow & B   &  :&  A \rangle .\\
                        & \langle i.2,&  B    & \rightarrow & I(A)&  :&  N_b \rangle .\\
                        & \langle i.3,&  I(A) & \rightarrow & B   &  :&  {Y} \rangle .\\
                        & \langle i.4,&  B    & \rightarrow & I(S)&  :& \{A.{Y} \}_{k_{bs}}\rangle .  \\
                        & \langle i.5,&  I(S) & \rightarrow & B&  :&  \{N_b^i.{Z}\}_{k_{bs}}\rangle \\
                    \end{array}
                    \end{array}\]
}
\normalsize
         The generalized role of $S$ is:
\small{
            \[\begin{array}{lllllll}
                {\cal   S}_G   ^1=   &  \langle  i.4,&  I(B)  &  \rightarrow  &  S  &  :&
                \{A,\{{U},{V}\}_{k_{as}} \}_{k_{bs}} \rangle . \\
                        & \langle i.5,&  S    & \rightarrow & I(B)&  :&  \{{U}.{V}\}_{k_{bs}}\rangle
                \end{array}\]
}
\normalsize
    Hence, the role-based specification  of  the  protocol  described  by Table~\ref{Wooo}  is  $  {\cal
    R}_G(p) = \{{\cal A}_G ^1,~{\cal A}_G ^2,~{\cal B}_G
    ^1,~{\cal B}_G ^2,~{\cal B}_G ^3,~{\cal S}_G ^1\}$. The role-based specification\index{Protocol!Role-Based Specification} is a  model to formalize  the cencept  of  valid  traces of a protocol. More details about the role-based specification are in~\cite{ContextWF,Debbabi11,Debbabi22,Debbabi33}. 
\enlargethispage{10mm}
\item[+] A valid trace is an instantiated messages of the generalized roles where each message sent by the intruder can be generated by her using her capacity and the prior messages. We denote by $ [\![p]\!]$ the set of valid traces of $p$.
\item[+] We denote by ${\cal{M}}_p^{\cal{G}}$ the set of messages with variables generated by $R_G(p)$, by ${\cal{M}}_p$ the set of closed messages generated by substituting all variables in ${\cal{M}}_p^{\cal{G}}$.  We denote by $R^+$ (resp. $R^-$) the set of sent messages (resp. received messages) by a honest agent in the role $R$. By convention, we use the uppercase letters for sequences or sets of elements and the lowercase for single elements. For example  $m$ denotes a single message, $M$ a set of messages, $R$ a role of composite steps, $r$ a step and $R.r$ a role ending by the step $r$.
\end{itemize}


%% file: Confpreuve.tex
\section{Secrecy in Increasing Protocols}\label{sectionPreuveThFond}

In this section, we give relaxed conditions allowing a function to be reliable for analysis. We prove that an increasing protocol is correct with respect to  secrecy  when analyzed with such functions.

\begin{defini}(Well-Formed Function)\label{bienforme}\index{Interpretation function}
$ $\\
Let $F$ be a function and ${\cal{C}}$ be a context of verification. 
$F$ is ${\cal{C}}$-well-formed  iff:\index{Fonction well-formed}\\
$\forall M,M_1,M_2 \subseteq {\cal{M}}, \forall \alpha \in {\cal{A}}({\cal{M}}) \mbox{:}\\$
\[
\left\{
    \begin{array}{lll}
         F(\alpha,\{\alpha\})&= & \bot \\
         F(\alpha,{M})& =&\top, \mbox{ if } \alpha \notin {\cal{A}}({M})  \\
         F(\alpha, {M}_1 \cup {M}_2)&= & F(\alpha, {M}_1)\sqcap F(\alpha,{M}_2) \\
    \end{array}
\right.
\]
\end{defini}

For an atom $\alpha$ in a set of messages $M$, a well-formed function returns the bottom value "$\bot$" if $M=\{\alpha\}$ (clear).  It returns the top value "$\top$" if it does not appear in this set. It returns for it in the union of two sets, the minimum "$\sqcap$" of the two values calculated in each set separately.

\begin{defini}(Full-invariant-by-intruder Function)\label{spi}\index{Stabilité par intrus}
$ $\\
Let $F$ be a function and ${\cal{C}}$ be a context of verification.  \index{Interpretation function}\\
$F$ is ${\cal{C}}$-full-invariant-by-intruder iff:\\
$
\forall {M} \subseteq {\cal{M}}, m\in {\cal{M}}. {M} \models_{\cal{C}} m \Rightarrow \forall \alpha \in {\cal{A}}(m). (F(\alpha,m) \sqsupseteq F(\alpha,{M}))  \vee (\ulcorner K(I) \urcorner \sqsupseteq \ulcorner \alpha \urcorner)
$
\end{defini}

A full-invariant-by-intruder $F$  should be such that it it attributes a security level  to a message $\alpha$ in $M$, the intruder can never produce from $M$ another message $m$ that decrease this level  (i.e. $F(\alpha,m) \sqsupseteq F(\alpha,{M})$) except when $\alpha$ is intended to be known by the intruder (i.e. $\ulcorner K(I) \urcorner \sqsupseteq \ulcorner \alpha \urcorner$).

\begin{defini}(Reliable  Function)\index{${\cal{C}}$-reliable}
$ $\\
Let $F$ be a function and ${\cal{C}}$ be a context of verification. $F  \mbox { is }{\cal{C}}\mbox{-reliable}  \mbox{ iff :}$\\
$F \mbox{ is } {\cal{C}-}\mbox{well-formed }\wedge F  \mbox{ is } {\cal{C}-}\mbox{full-invariant-by-intruder}.$
\end{defini} 

A reliable function is simply a function that is well-formed and full-invariant-by-intruder in a given context of verification ${\cal{C}}$.

\begin{defini}($F$-Increasing Protocol)\label{ProAbsCroiArticle}
$ $\\
Let $F$ be a function, ${\cal{C}}$ be a context of verification  and $p$ be a protocol. 
$p$ is $F$-increasing in ${\cal{C}}$ if:\\
$\forall R.r \in R_G(p),\forall \sigma \in \Gamma: {\cal{X}} \rightarrow {\cal{M}}_p  \mbox{ we have: }$
\[
\forall \alpha \in {\cal{A}}({\cal{M}}).F(\alpha, r^+\sigma)\sqsupseteq \ulcorner \alpha \urcorner \sqcap F(\alpha, R^-\sigma)
\]
\end{defini}

A $F$-increasing protocol produces  valid traces (interleaving of substituted messages in generalized roles) where every involved principal (every substituted generalized role) never decreases the security levels, calculated by $F$, of received components. 

\begin{defini}(Secret Disclosure)\label{divulgationArticle}\index{Divulgation}
$ $\\
Let $p$ be a protocol and ${\cal{C}}$ be a context of verification.\\
We say that $p$ discloses a secret $\alpha\in {\cal{A}}({\cal{M}})$  in ${{\cal{C}}}$ if:
\[\exists \rho\in [\![p]\!].(\rho \models_{{\cal{C}}} \alpha)\wedge(\ulcorner K(I) \urcorner \not \sqsupseteq \ulcorner \alpha \urcorner)\]
\end{defini}

A secret disclosure consists in manipulating a valid trace $\rho$ by the intruder, using her knowledge $K(I) $ in a context of verification ${{\cal{C}}}$, to infer a secret $\alpha$ that she is not intended to know ($\ulcorner K(I) \urcorner \not \sqsupseteq \ulcorner \alpha \urcorner$).

\begin{lem}\label{lemprincArticle}
$ $\\
Let $F$ be a ${\cal{C}}$-reliable function and $p$ a $F$-increasing protocol. We have: $\forall m \in {\cal{M}}. [\![p]\!] \models_{\cal{C}} m \Rightarrow$
$
\forall \alpha \in {\cal{A}}(m). (F(\alpha,m)\sqsupseteq \ulcorner \alpha \urcorner)\vee (\ulcorner K(I) \urcorner \sqsupseteq \ulcorner \alpha \urcorner)
$
\end{lem}

\noindent\textit{See the proof \ref{PR4} in ~\cite{preuvesJ}}.\\
\vspace*{-0.1\baselineskip}
The lemma \ref{lemprincArticle} states that for any atom $\alpha$ in a message generated by an increasing protocol, its security level calculated by a reliable function is always greater than its initial value given in the context, provided that the intruder is not initially allowed to know it. Indeed, initially the atom has a some security level. This level cannot be decreased by the intruder using her initial knowledge and received messages since a reliable function is full-invariant-by-intruder. In each new step of the evolution of the valid trace, involved messages are better protected since the protocol is increasing.  The proof is run by induction on the size (evolution) of the trace and uses the properties of reliability of the function in every new step of the induction.

\begin{thm}(Security of Increasing Protocols)\label{mainThArticle}
$ $\\
Let $F$ be a ${\cal{C}}$-reliable function and $p$ a $F$-increasing protocol.
\begin{center}
$p$ is ${\cal{C}}$-correct with respect to secrecy.
\end{center}\index{Confidentialité}

\end{thm}

\rhead{}
\lhead{}
\lfoot{\textit{\\}\includegraphics{logo3.png} }
\cfoot{}

\begin{preuve2}\label{PR5Article} 
$ $\\
Let's suppose that $p$ reveals some atomic secret $\alpha$. \\
From the definition \ref{divulgationArticle} we have:
\begin{equation}\exists \rho\in [\![p]\!].(\rho \models_{\cal{C}} \alpha)\wedge(\ulcorner K(I) \urcorner \not \sqsupseteq \ulcorner \alpha \urcorner) \label{preuvthe1Article}\end{equation}
Since $F$ is ${\cal{C}}$-reliable and $p$ is an $F$-increasing protocol, we have from the lemma \ref{lemprincArticle}:
\begin{equation}(F(\alpha,\alpha)\sqsupseteq \ulcorner \alpha \urcorner)\vee (\ulcorner K(I) \urcorner \sqsupseteq \ulcorner \alpha \urcorner)\label{preuvthe2Article}\end{equation}
From (\ref{preuvthe1Article}) and (\ref{preuvthe2Article}) we deduce:
\begin{equation}F(\alpha,\alpha)\sqsupseteq \ulcorner \alpha \urcorner \label{preuvthe3Article}\end{equation}
As $F$ is ${\cal{C}}$-well-formed, we have:
\begin{equation}F(\alpha,\alpha)= \bot \label{preuvthe4Article}\end{equation}
From (\ref{preuvthe3Article}) and (\ref{preuvthe4Article}) we deduce:
\begin{equation}\bot= \ulcorner \alpha \urcorner \label{preuvthe5corArticle}\end{equation}
(\ref{preuvthe5corArticle}) is not possible because it is opposite to:  $\ulcorner K(I) \urcorner \not \sqsupseteq \ulcorner \alpha \urcorner$ in (\ref{preuvthe1Article}).\\
\\
Conclusion: $p$ is ${\cal{C}}$-correct with respect to secrecy.
\end{preuve2} 

%% file: ConclusionLeger.tex
\section{Comparison with Related Works}
The theorem \ref{mainThArticle} states that an increasing protocol is correct with respect to secrecy when analyzed with a function that is  well-formed and  full-invariant by intruder, or shortly reliable. Compared to the conditions imposed by Houmani et al. in~\cite{Houmani1,Houmani5} to their functions, we have one condition less. Indeed, Houmani et al. expect from  a protocol to be increasing on the messages of generalized roles whatever the substitutions they may receive (when the protocol is running) and demanded from the interpretation function to resist to these substitutions. As a result, even if they gave a guideline to build reliable functions, just two functions could be really built: the function DEK and the function DEKAN. We notice as well that the function DEK is disappointing in practice. That is mainly due to the complexity to  prove that a function satisfies the full-invariance by substitution property. In this paper, we release our functions from this hard condition in order to be more comfortable in building functions. We rehouse this condition in our new definition of an increasing protocol. The problem of substitution goes to the protocol and becomes relatively less difficult to handle. Our approach has in common with  Schneider's method and Houmani's method the idea of transforming the problem of secrecy to a problem of protocol growth. All of these three approaches decide only if the protocol is proved increasing using a reliable function, as well. 
\section{Conclusion and Future Work}
{\normalsize
Releasing a function from a condition may motivate us to undertake special cautions when using it. In a future work, we introduce the notion of witness-functions to analyze cryptographic protocols. A witness-function is a protocol-dependent function. It introduces new derivation techniques to solve the problem of substitution locally in the protocol. It 	
supplies two bounds that are independent of all substitutions which enables any decision made on the generalized roles to be sent to valid traces. This replaces the restrictive condition of full-invariance by substitution in Houmani's logic. The witness-functions were successful to prove secrecy in many protocols. They could even tell about flaws.}

%% file: Appendix.tex
\rhead{}
\lhead{}
\lfoot{}
\cfoot{}
\begin{center}
{\Large{RELAXED CONDITIONS FOR SECRECY IN CRYPTOGRAPHIC PROTOCOLS: PROOFS AND INTERMEDIATE RESULTS}}
\end{center}
\begin{center}
$\mbox{by: Jaouhar Fattahi, Mohamed Mejri and Hanane Houmani- April 2014}$
\end{center}
$ $\\
$ $\\
\subsection*{Valid trace}
Before defining a valid trace, let's give some intermediate definitions.\\

1- Step: it is defined by the following BNF grammar:
\[step :: \langle j.i, A \longrightarrow I (B) : m\rangle  | \langle j.i, I  (a) \longrightarrow B : m\rangle
\]

where $j$ is the identifier of the session and $i$ is the identifier of the step.\\

2- Trace: a trace $\rho$ is a sequence of steps. It is defined by the following BNF grammar:

\[\rho :: \varepsilon \mbox{ }| \mbox{ } step \mbox{ }|\mbox{ } \rho.step
\]

where $\varepsilon$ is the empty trace. We denote by $\bar{\rho}$ all the steps of $\rho$.\\

3- Session identifier: every step has an identifier. The set of identifiers ${\cal{S}}^{\rho}$ associated with a trace $\rho$ is built as follows:
\begin{center}
\begin{tabular}{r c l}
   ${\cal{S}}^{\varepsilon}$& $=$ & $\emptyset$ \\
   ${\cal{S}}^{\rho.\langle j.i,A\longrightarrow I(B):m\rangle }$& $=$ & ${\cal{S}^{\rho}}\cup \{j\}$ \\
   ${\cal{S}}^{\rho.\langle j.i,I (a)\longrightarrow B:m\rangle }$& $=$ & ${\cal{S}^{\rho}}\cup\{j\}$ \\
\end{tabular}
\end{center}

4- Session: a trace is an interleaving of many sessions where each of them has an identifier. We define a session (identified by ${Id}$) associated with a trace $\rho^n$ (coming from an interleaving of many sessions) as follows:

\begin{center}
\begin{tabular}{r c l c}
   ${\varepsilon}^{Id}$& $=$ & ${\varepsilon}$ \\
   $(\rho.\langle j.i,A\longrightarrow I(B):m\rangle )^{Id}$& $=$ & $\rho^{Id}$ &\mbox{      if } ${Id}\neq j$\\
   $(\rho.\langle j.i,I (a)\longrightarrow B:m\rangle )^{Id}$& $=$ & $\rho^{Id}$ &\mbox{      if } ${Id}\neq j$\\
   $(\rho.\langle {Id}.i,A\longrightarrow I(B):m\rangle )^{Id}$& $=$ & $\rho^{Id}.\langle {Id}.i,A\longrightarrow I(B):m\rangle $ \\
   $(\rho.\langle {Id}.i,I (a)\longrightarrow B:m\rangle )^{Id}$& $=$ & $\rho^{Id}.\langle {Id}.i,I (a)\longrightarrow B:m)\rangle $ \\

\end{tabular}
\end{center}

where $\varepsilon$ is the empty trace. It may be eliminated when preceded or followed by a non-empty trace.\\

5- $\mbox{Def}/\mbox{Use}$: the knowledge of the intruder change from a step to another. After an execution of  a trace, the intruder acquires new knowledge or defines herself news messages. We denote by $\mbox{Def}(\rho)$ the knowledge (messages) received by the intruder after execution of a trace $\rho$, and by $\mbox{Use}(\rho)$ the messages she builds after the execution of $\rho$. $\mbox{Use}(\rho)$ and $\mbox{Def}(\rho)$ are defined as follows:

\begin{center}
\begin{tabular}{r c l}
   $\mbox{Def}({\varepsilon})$& $=$ & $\emptyset$ \\
   $\mbox{Def}({\rho.\langle j.i,A\longrightarrow I(B):m\rangle })$& $=$ & $\mbox{Def}(\rho)\cup \{m\} $ \\
   $\mbox{Def}({\rho.\langle j.i,I (a)\longrightarrow B:m\rangle })$& $=$ & $\mbox{Def}(\rho)$ \\
\end{tabular}
\end{center}

\begin{center}
\begin{tabular}{r c l}
   $\mbox{Use}({\varepsilon})$& $=$ & $\emptyset$ \\
   $\mbox{Use}({\rho.\langle j.i,A\longrightarrow I(B):m\rangle })$& $=$ & $\mbox{Use}(\rho)$ \\
   $\mbox{Use}({\rho.\langle j.i,I (a)\longrightarrow B:m\rangle })$& $=$ & $\mbox{Use}(\rho)\cup \{m\} $  \\
\end{tabular}
\end{center}

For the sake of simplification, we denote by:
\begin{center} $\mbox{Use}(\rho)=\rho^+$ and $\mbox{Def}(\rho)=\rho^-$ \end{center}

6- ${\cal{C}}$-well-defined trace:
a trace is said ${\cal{C}}$-well-defined when the intruder is able to define all the messages contained in the trace before sending them. i.e for $\rho=\rho_1.e.\rho_2$ we have:
\[
{\rho_1}^+\models_{K(I)} e^-
\]
where $e$ is a step of communication and ${K(I)}$ are the knowledge of the intruder.\\

7- ${\cal{C}}$-well-formed trace: a trace is said ${\cal{C}}$-well-formed when it is generated by substitution in a generalized role of the protocol. i.e. $\exists r \in R_G(p)$, a session $i\in {\cal{S}}^\rho$ and a substitution $\sigma\in \Gamma$ (the set of all substitutions) such that:
\[
{\rho}^{i}= r\sigma
\]

8- Valid trace: $\rho$ is a valide trace of a protocol $p$ when $\rho$ is ${\cal{C}}$-well-defined and $\rho$ is ${\cal{C}}$-well-formed.\\

A valide trace is a run of a protocol. It is an instance of the generalized roles produced by the intruder (or a regular agent) with respect to the rules of the protocol and the context of verification. 

\begin{defini}(Prefix of a trace)\label{PrefixeTrace}
$ $\\
Let $\rho_1$ and $\rho_2$ be two traces.
\begin{center}
$\rho_2$ is a prefix of $\rho_1$ if $\exists \rho_3$ such that: $\rho_1=\rho_2.\rho_3$
\end{center}
\end{defini}
\begin{defini}(Size of a trace)\label{TailleTrace}
$ $\\
The size of a trace $\rho$ denoted by $|\rho|$ is defined as follows:
\begin{center}
\begin{tabular}{rcl}
  $|{\varepsilon}|$ & $=$ & $0$ \\
  $|step|$ & $=$ & $1$ \\
  $|\rho.\rho'|$ & $=$ & $|\rho|+|\rho'|$  \\
\end{tabular}
\end{center}
\end{defini} 


\begin{defini}(Well-formed  function)\label{bienformeProofs}
$ $\\
Let $F$ be a function and ${\cal{C}}$ a context of verification. \\
$F$ is well-formed in ${\cal{C}}$ if:\index{Fonction well-formed}\\
$\forall M,M_1,M_2 \subseteq {\cal{M}}, \forall \alpha \in {\cal{A}}({\cal{M}}) \mbox{:}\\$
\[
\left\{
    \begin{array}{lll}
         F(\alpha,\{\alpha\})&= & \bot \\
         F(\alpha, {M}_1 \cup {M}_2)&= & F(\alpha, {M}_1)\sqcap F(\alpha,{M}_2) \\
         F(\alpha,{M})& =&\top, \mbox{ if } \alpha \notin {\cal{A}}({M})  \\
    \end{array}
\right.
\]
\end{defini}

For an atom $\alpha$ in a set of messages $M$, a well-formed function returns the bottom value "$\bot$", if $M=\{\alpha\}$. It returns for it in the union of two sets, the minimum "$\sqcap$" of the two values calculated in each set separately. It returns the top value "$\top$", if it does not appear in this set.

\begin{defini}(Full-invariant-by-intruder function)\label{spiProofs}
$ $\\
Let $F$ be a function and ${\cal{C}}$ a context of verification.  \index{Interpretation function}\\
$F$ is full-invariant-by-intruder in ${\cal{C}}$ if:\\
$
\forall {M} \subseteq {\cal{M}}, m\in {\cal{M}}. {M} \models_{\cal{C}} m \Rightarrow \forall \alpha \in {\cal{A}}(m). (F(\alpha,m) \sqsupseteq F(\alpha,{M}))  \vee (\ulcorner K(I) \urcorner \sqsupseteq \ulcorner \alpha \urcorner)
$
\end{defini}

A reliable function $F$ should be full-invariant-by-intruder. That is to say, if $F$ attributes a security level  to a message $\alpha$ in $M$, then the intruder can never produce from $M$ another message $m$ that decrease this level  (i.e. $F(\alpha,m) \sqsupseteq F(\alpha,{M})$) except when $\alpha$ is intended to be known by the intruder (i.e. $\ulcorner K(I) \urcorner \sqsupseteq \ulcorner \alpha \urcorner$).

\begin{defini}(Reliable  function)\label{ReliableFunctionsProofs}
$ $\\
Let $F$ be a function and ${\cal{C}}$ a context of verification.
\[F  \mbox { is }{\cal{C}}\mbox{-reliable}  \mbox{ if } F \mbox{ is well-formed and } F \mbox{ is full-invariant-by-intruder in } {\cal{C}}.\]
\end{defini} 

A reliable function is simply a function that is well-formed and full-invariant-by-intruder in a given context of verification ${\cal{C}}$.

\begin{defini}($F$-increasing protocol)\label{ProAbsCroiProofs}
$ $\\
Let $F$ be a function, ${\cal{C}}$ a context of verification  and $p$ a protocol.\\
$p$ is $F$-increasing in ${\cal{C}}$ if:\\
$\forall R.r \in R_G(p),\forall \sigma \in \Gamma: {\cal{X}} \rightarrow {\cal{M}}_p  \mbox{ we have: }$
\[
\forall \alpha \in {\cal{A}}({\cal{M}}).F(\alpha, r^+\sigma)\sqsupseteq \ulcorner \alpha \urcorner \sqcap F(\alpha, R^-\sigma)
\]
\end{defini}

A $F$-increasing protocol produces  valid traces (interleaving of substituted generalized roles) where every involved principal (every substituted generalized role) never decreases the security levels of received components. When a protocol is  $F$-increasing and $F$ is a reliable function, it will be easy to prove its correctness with respect to the secrecy property. In fact, if every agent appropriately  protects her sent messages (if she initially knows the security level of a component, she has to encrypt it with at least one key having a similar or higher security level, and if she does not know its security level,  she estimates it using a reliable function), the intruder can never reveal it.

\begin{defini}(Secret disclosure)\label{divulgationProofs}
$ $\\
Let $p$ be a protocol and ${\cal{C}}$ a context of verification.\\
We say that $p$ discloses a secret $\alpha\in {\cal{A}}({\cal{M}})$  in ${{\cal{C}}}$ if:
\[\exists \rho\in [\![p]\!].(\rho \models_{{\cal{C}}} \alpha)\wedge(\ulcorner K(I) \urcorner \not \sqsupseteq \ulcorner \alpha \urcorner)\]
\end{defini}

A secret disclosure consists in exploiting a valid trace of the protocol (denoted by $ [\![p]\!]$) by the intruder using her knowledge $K(I) $ in a context of verification ${{\cal{C}}}$, to infer a secret $\alpha$ that she is not allowed to know (expressed by: $\ulcorner K(I) \urcorner \not \sqsupseteq \ulcorner \alpha \urcorner$).

\begin{defini}(${\cal{C}}$-correct protocol with respect to the secrecy property)\label{correctionProofs}
$ $\\
Let $p$ be a protocol and ${\cal{C}}$ a context of verification. \\
$p$ is ${\cal{C}}$-correct with respect to the secrecy property  if:
\[
\forall \alpha\in {\cal{A}}({\cal{M}}), \forall \rho \in [\![p]\!]. \rho \models_{{\cal{C}}} \alpha \Rightarrow \ulcorner K(I) \urcorner  \sqsupseteq \ulcorner \alpha \urcorner
\]
We denote:
\[\forall \alpha\in {\cal{A}}({\cal{M}}). [\![p]\!] \models_{\cal{C}}  \alpha \Rightarrow \ulcorner K(I) \urcorner  \sqsupseteq \ulcorner \alpha \urcorner
\]

\end{defini}

A correct protocol is such that any valid trace $\rho$ that it produces never leaks a secret that the intruder is not allowed to know. 

\begin{defini}($\sqsupseteq_F$)\label{RelatSupF}
$ $\\
$ $\\
Let ${\cal{C}}=\langle{\cal{M}},\xi,\models,{\cal{K}},{\cal{L}}^\sqsupseteq,\ulcorner .\urcorner\rangle$ be a verification context.\\
Let $F$ be a function. \index{Interpretation function}\\
Let $M_1$, $M_2$ $\subseteq$ ${\cal{M}}$.\\
\[
M_1 \sqsupseteq_F M_2 \Longleftrightarrow \forall \alpha \in {\cal{A}}(M_1).F(\alpha,M_1) \sqsupseteq F(\alpha,{M_2})
\]

\end{defini}

\begin{defini}(Auxiliary function)\label{bienformederivee}\index{Interpretation function}
$ $\\
$ $\\
Let ${\cal{C}}=\langle{\cal{M}},\xi,\models,{\cal{K}},{\cal{L}}^\sqsupseteq,\ulcorner .\urcorner\rangle$ be a verification context.\\
Let $F$ be a function.\\
We define the auxiliary function of $F$, denoted by $\hat{F}$, as follows:
\[
\begin{array}{cccc}
  \hat{F}: & {\cal{A}}\times {\cal{M}} & \longmapsto & {\cal{L}}^\sqsupseteq \\
   & (\alpha,m) & \longmapsto & \ulcorner\alpha\urcorner \sqcap F(\alpha,m)
\end{array}
\]

\end{defini}

\begin{defini}($\sqsupseteq_{\hat{F}}$)\label{RelatSupHatF}
$ $\\
$ $\\
Let ${\cal{C}}=\langle{\cal{M}},\xi,\models,{\cal{K}},{\cal{L}}^\sqsupseteq,\ulcorner .\urcorner\rangle$ be a verification context.\\
Let $F$ be a function. \index{Interpretation function}\\
Let $M_1$, $M_2$ $\subseteq$ ${\cal{M}}$\\
\[
M_1 \sqsupseteq_{\hat{F}} M_2 \Longleftrightarrow \forall \alpha \in {\cal{A}}(M_1).{\hat{F}}(\alpha,M_1) \sqsupseteq {\hat{F}}(\alpha,{M_2})
\]
\end{defini}

\begin{pre}
$ $\\
$ $\\
Let ${\cal{C}}=\langle{\cal{M}},\xi,\models,{\cal{K}},{\cal{L}}^\sqsupseteq,\ulcorner .\urcorner\rangle$ be a verification context.\\
Let $F$ be a function.\\
Let $\hat{F}$ be the auxiliary function of $F$.\\
\\
If $F$ is well-formed in ${\cal{C}}$ then:
\\
\\
$\forall M,M_1,M_2 \subseteq {\cal{M}}, \forall \alpha \in {\cal{A}}({\cal{M}}).\\$
\[
\left\{
    \begin{array}{lll}
         \hat{F}(\alpha,\{\alpha\})&= & \bot \\
         &\mbox{and} &\\
         \hat{F}(\alpha, {M}_1 \cup {M}_2)&= & \hat{F}(\alpha, {M}_1)\sqcap \hat{F}(\alpha,{M}_2) \\
         &\mbox{and} &\\
         \hat{F}(\alpha,{M})& =&\ulcorner\alpha\urcorner, \mbox{ if } \alpha \notin {\cal{A}}({M})  \\
    \end{array}
\right.
\]

\end{pre}
$ $\\
See the proof \ref{PR1}.

\begin{preuve}\label{PR1}
$ $\\
$ $\\
$F$ is well-formed in ${\cal{C}}$ then we have from the definition \label{bienformeProofs}:
\\
\\
$\forall M,M_1,M_2 \subseteq {\cal{M}}, \forall \alpha \in {\cal{A}}({\cal{M}}).\\$
\begin{equation}
\left\{
    \begin{array}{lll}
         F(\alpha,\{\alpha\})&= & \bot \\
         &\mbox{and} &\\
         F(\alpha, {M}_1 \cup {M}_2)&= & F(\alpha, {M}_1)\sqcap F(\alpha,{M}_2) \\
         &\mbox{and} &\\
         F(\alpha,{M})& =&\top, \mbox{ if } \alpha \notin {\cal{A}}({M})  \\
    \end{array}
\right.\label{eq1bienformee}
\end{equation}
From \ref{eq1bienformee}, we have the three following results:
\begin{itemize}
  \item [$\bullet$]\begin{equation}F(\alpha,\{\alpha\})=\bot \label{eq2bienformee}\end{equation}
  From \ref{eq2bienformee} and since ${\cal{L}}^\sqsupseteq$ is a lattice, we have:
  \begin{equation}\ulcorner\alpha\urcorner \sqcap F(\alpha,\{\alpha\})=\bot \label{biseq2bienformee}\end{equation}
  From \ref{biseq2bienformee} and the definition \ref{bienformederivee} of $\hat{F}$ we have:
  \begin{equation}\hat{F}(\alpha,\{\alpha\})=\bot \label{biseq3bienformee}\end{equation}
  \item [$\bullet$]\begin{equation}F(\alpha, {M}_1 \cup {M}_2)=F(\alpha, {M}_1)\sqcap F(\alpha,{M}_2)\label{biseq4bienformee}\end{equation}
  From \ref{biseq4bienformee} and since ${\cal{L}}^\sqsupseteq$ is a lattice we have:
    \begin{equation}\ulcorner\alpha\urcorner \sqcap F(\alpha, {M}_1 \cup {M}_2)=\ulcorner\alpha\urcorner\sqcap (F(\alpha, {M}_1)\sqcap F(\alpha,{M}_2))\label{biseq5bienformeeModi}\end{equation}
  Or else:
  \begin{equation}\ulcorner\alpha\urcorner \sqcap F(\alpha, {M}_1 \cup {M}_2)=\ulcorner\alpha\urcorner\sqcap F(\alpha, {M}_1)\sqcap \ulcorner\alpha\urcorner\sqcap F(\alpha,{M}_2)\label{biseq5bienformee}\end{equation}

  From \ref{biseq5bienformee} and from the definition \ref{bienformederivee} of $\hat{F}$ we have:
  \begin{equation}\hat{F}(\alpha, {M}_1 \cup {M}_2)=\hat{F}(\alpha, {M}_1)\sqcap \hat{F}(\alpha,{M}_2)\label{biseq6bienformee}\end{equation}

  \item [$\bullet$]\begin{equation}F(\alpha,{M})=\top \mbox{, if } \alpha \notin {M}\label{eq4bienformee}\end{equation}
  From \ref{eq4bienformee} and since ${\cal{L}}^\sqsupseteq$ is a lattice and $\ulcorner\alpha\urcorner \sqsubseteq \top$ we have:
      \begin{equation}\ulcorner\alpha\urcorner\sqcap F(\alpha,{M})=\ulcorner\alpha\urcorner \mbox{, if } \alpha \notin {\cal{A}}(M)\label{biseq7bienformee}\end{equation}
  From \ref{biseq7bienformee} and from the definition \ref{bienformederivee} of $\hat{F}$ we have:
  \begin{equation}\hat{F}(\alpha,{M})=\ulcorner\alpha\urcorner \mbox{, if } \alpha \notin {\cal{A}}(M)\label{biseq8bienformee}\end{equation}
\end{itemize}
From \ref{biseq3bienformee}, \ref{biseq6bienformee} and \ref{biseq8bienformee} we have:\\
\\
$\forall M,M_1,M_2 \subseteq {\cal{M}}, \forall \alpha \in {\cal{A}}({\cal{M}}).\\$
\[
\left\{
    \begin{array}{lll}
         \hat{F}(\alpha,\{\alpha\}&= & \bot \\
         &\mbox{and} &\\
         \hat{F}(\alpha, {M}_1 \cup {M}_2)&= & \hat{F}(\alpha, {M}_1)\sqcap \hat{F}(\alpha,{M}_2) \\
         &\mbox{and} &\\
         \hat{F}(\alpha,{M})& =&\ulcorner\alpha\urcorner, \mbox{ if } \alpha \notin {\cal{A}}({M})  \\
    \end{array}
\right.
\]
\end{preuve}

\begin{pre}
$ $\\
$ $\\
Let ${\cal{C}}=\langle{\cal{M}},\xi,\models,{\cal{K}},{\cal{L}}^\sqsupseteq,\ulcorner .\urcorner\rangle$ be a verification context.\\
Let $F$ be a function.\\
Let $\hat{F}$ the auxiliary function of $F$.\\
If $F$ is full-invariant-by-intruder in ${\cal{C}}$ then $\hat{F}$ is full-invariant-by-intruder in ${\cal{C}}$.
\end{pre}
$ $\\
See the proof \ref{PR2}.

\begin{preuve}\label{PR2}
$ $\\
$ $\\
Let ${M} \subseteq {\cal{M}}$ and $m \in {\cal{M}}$ such that:
\begin{equation}
 {M} \models_{{\cal{C}}} m
\label{eq1spihatModif}
\end{equation}
From the definition \ref{spiProofs} and \ref{eq1spihatModif} we have for all $\alpha\in {\cal{A}}(m)$:\\
\begin{equation}
(F(\alpha,m) \sqsupseteq F(\alpha,{M}))  \vee (\ulcorner K(I) \urcorner \sqsupseteq \ulcorner \alpha \urcorner)\label{eq1spihat}
\end{equation}
From \ref{eq1spihat} we have:\\
\begin{itemize}
  \item [$\bullet$]either: \begin{equation}F(\alpha,m) \sqsupseteq F(\alpha,{M})\label{eq3spihat}
\end{equation}
From \ref{eq3spihat} and since ${\cal{L}}^\sqsupseteq$ is a lattice, we have:\\
 \begin{equation}\ulcorner \alpha \urcorner \sqcap F(\alpha,m) \sqsupseteq \ulcorner \alpha \urcorner \sqcap F(\alpha,{M})\label{eq5spihat}
\end{equation}
From \ref{eq5spihat} and from the definition \ref{bienformederivee} de $\hat{F}$ we have:
 \begin{equation}\hat{F}(\alpha,m) \sqsupseteq \hat{F}(\alpha,{M})\label{eq6spihat}
\end{equation}
  \item [$\bullet$] or: \begin{equation}\ulcorner K(I) \urcorner \sqsupseteq \ulcorner \alpha \urcorner\label{eq7spihat}
\end{equation}
\end{itemize}
From \ref{eq6spihat} and \ref{eq7spihat} we have for all ${M} \subseteq {\cal{M}}$ and $m \in {\cal{M}}$:\\
\[
 {M} \models_{{\cal{C}}} m \Rightarrow \forall \alpha \in {\cal{A}}(m). (\hat{F}(\alpha,m) \sqsupseteq \hat{F}(\alpha,{M}))  \vee (\ulcorner K(I) \urcorner \sqsupseteq \ulcorner \alpha \urcorner)
\]
Then $\hat{F}$ is full-invariant-by-intruder in ${\cal{C}}$.
\end{preuve}

\begin{lem}\label{lem411Proofs}
$ $\\
$ $\\
Let ${\cal{C}}=\langle{\cal{M}},\xi,\models,{\cal{K}},{\cal{L}}^\sqsupseteq,\ulcorner .\urcorner\rangle$ be a verification context.\\
Let $F$ be a ${\cal{C}}$-reliable function.\\
Let $p$ be a protocol.\\
\\
We have:\\

If ($p$ is $F$-increasing in ${\cal{C}}$)$ \mbox{ then } (\forall \rho.e \in [\![p]\!]\mbox{ we have: }e^+\sqsupseteq_{\hat{F}} \rho^-)$

\end{lem}
$ $\\
See the proof \ref{PR3}.

\begin{preuve}\label{PR3}
$ $\\
$ $\\
Since $p$ is an $F$-increasing protocol in ${\cal{C}}$, then from the definition \ref{ProAbsCroiProofs} we have:\\
\\
$\forall R.r \in R_G(p), \forall \sigma\in \Gamma: X \rightarrow {\cal{M}}_p(closed)$ we have:\\
\begin{equation}\forall \alpha \in {{\cal{A}}(r^+\sigma)}.F(\alpha, r^+\sigma)\sqsupseteq \ulcorner \alpha \urcorner \sqcap F(\alpha, R^-\sigma)\label{eq1trace}\end{equation} \\
Since $\rho.e$ is a valid trace, then exists a substitution $\sigma\in \Gamma$, a generalized role $R.r$ $\in R_G(p)$ and a session $i=S(e)$ such that:
\begin{equation}(R\sigma=\rho^i)\wedge(r\sigma=e) \label{eq2trace}\end{equation}
From \ref{eq1trace} and \ref{eq2trace}, we have:
\begin{equation}\forall \alpha \in {\cal{A}}(e^+).F(\alpha,e^+)\sqsupseteq \ulcorner \alpha \urcorner \sqcap F(\alpha,\rho^{i-})\label{eq3trace}\end{equation}
Since ${\cal{L}}^\sqsupseteq$ is a lattice, we have:
\begin{equation}\forall \alpha \in {\cal{A}}(e^+).\ulcorner \alpha \urcorner \sqcap F(\alpha,e^+)\sqsupseteq \ulcorner \alpha \urcorner \sqcap \ulcorner \alpha \urcorner \sqcap F(\alpha,\rho^{i-})\label{eq4trace}\end{equation}
From \ref{eq4trace} and since ${\cal{L}}^\sqsupseteq$ is a lattice we have:
\begin{equation}\forall \alpha \in {\cal{A}}(e^+).\ulcorner \alpha \urcorner \sqcap F(\alpha,e^+)\sqsupseteq \ulcorner \alpha \urcorner \sqcap F(\alpha,\rho^{i-})\label{eq5trace}\end{equation}
From the definition of $\hat{F}$ and from \ref{eq5trace}, we have:
\begin{equation}\forall \alpha \in {\cal{A}}(e^+). \hat{F}(\alpha,e^+)\sqsupseteq \hat{F}(\alpha,\rho^{i-}) \label{eq6trace}\end{equation}
From \ref{eq6trace} and the definition of $\sqsupseteq_{\hat{F}}$, we have:
\begin{equation}e^+\sqsupseteq_{\hat{F}}\rho^{i-}\label{eq7trace}\end{equation}
Since $\mbox{Use}(\rho^{i})\subseteq \mbox{Use}(\rho) $, from the definition of $\sqsupseteq_{\hat{F}}$ and since ${\cal{L}}^\sqsupseteq$ is a lattice, then we have:
\begin{equation}\rho^{i-} \sqsupseteq_{\hat{F}} \rho^-\label{eq8trace}\end{equation}
From \ref{eq7trace} and \ref{eq8trace}, we deduce that:
 \[e^+ \sqsupseteq_{\hat{F}} \rho^-\]

\end{preuve}

\begin{lem}\label{lemprincProofs}
$ $\\
Let $F$ be a ${\cal{C}}$-reliable function and $p$ a $F$-increasing protocol.\\
We have:
\[
\forall m \in {\cal{M}}. [\![p]\!] \models_{\cal{C}} m \Rightarrow \forall \alpha \in {\cal{A}}(m). (F(\alpha,m)\sqsupseteq \ulcorner \alpha \urcorner)\vee (\ulcorner K(I) \urcorner \sqsupseteq \ulcorner \alpha \urcorner).
\]

\end{lem}

The lemma \ref{lemprincProofs} asserts that for any atom $\alpha$ in a message generated by an increasing protocol, its security level calculated by a reliable function is maintained greater than its initial value in the context, if the intruder is not initially allowed to know it. Thus, initially the atom has a certain security level. This value cannot be decreased by the intruder using her initial knowledge and received messages since reliable functions are full-invariant-by-intruder. In each new step of any valid trace, involved messages are better protected since the protocol is increasing.  The proof is run by induction on the size of the trace and uses the reliability properties of the function in every step.\\
$ $\\
\\
See the proof \ref{PR4}

\begin{preuve}\label{PR4}
$ $\\
$ $\\
Let $\rho\in [\![p]\!]$ and $m \in {\cal{M}}$ such that: \begin{equation}\rho^+ \models_{\cal{C}} m \label{eq1preuvesti}\end{equation}
Let's prove the lemma by induction on the size of the trace.\\
\begin{itemize}
\item [$\bullet$] $|\rho|$=1.\\
\\
In this case, we have two sub-cases:\\
\begin{itemize}
\item[$\bullet$] either $\rho^+=\emptyset$, and so we have:\\
\\
From \ref{eq1preuvesti}, we have: \begin{equation}\emptyset \models_{\cal{C}} m\label{eq2preuvesti}\end{equation}
Since $F$ is full-invariant-by-intruder in ${\cal{C}}$, we have:
\begin{equation}\forall \alpha \in {\cal{A}}(m).(F(\alpha,m) \sqsupseteq F(\alpha,\emptyset))  \vee (\ulcorner K(I) \urcorner \sqsupseteq \ulcorner \alpha \urcorner) \label{eq3preuvesti}\end{equation}
Since $\alpha \notin \emptyset$ and $F$ is well-formed in ${\cal{C}}$, we have: \begin{equation}F(\alpha,\emptyset)=\top \sqsupseteq \ulcorner \alpha \urcorner\label{eq4preuvesti}\end{equation}
From \ref{eq3preuvesti}  and \ref{eq4preuvesti}, we have:
\begin{equation}\forall \alpha \in {\cal{A}}(m).(F(\alpha,m) \sqsupseteq \ulcorner \alpha \urcorner)  \vee (\ulcorner K(I) \urcorner \sqsupseteq \ulcorner \alpha \urcorner) \label{eq5preuvesti}\end{equation}
\end{itemize}
\begin{itemize}
\item[$\bullet$] or $\rho^+\neq\emptyset$, and so we have:
\newline
\newline
Since $|\rho|$=1 then exists an empty trace $\varepsilon$ such that $\rho=\varepsilon\rho^+$
\newline
\\
Since $F$ is full-invariant-by-intruder in ${\cal{C}}$, we have:\\
\begin{equation}\forall \alpha \in {\cal{A}}(m).(F(\alpha,m) \sqsupseteq F(\alpha,\rho^+))  \vee (\ulcorner K(I) \urcorner \sqsupseteq \ulcorner \alpha \urcorner)\label{eq6preuvesti}\end{equation}
Since $p$ is $F$-increasing in ${\cal{C}}$, then from the lemma~\ref{lem411Proofs}, we have:
\begin{equation}\rho^+\sqsupseteq_{\hat{F}} \varepsilon^-\label{eq7preuvesti}\end{equation}
Or else: \begin{equation}\forall \alpha \in {\cal{A}}(\rho^+).\hat{F}(\alpha,\rho^+) \sqsupseteq
\hat{F}(\alpha,\varepsilon^-)\label{eq8preuvestibis}\end{equation}
\newline
From \ref{eq8preuvestibis} and from the definition of $\hat{F}$ we have: \begin{equation}\forall \alpha \in {\cal{A}}(\rho^+).F(\alpha,\rho^+) \sqcap \ulcorner \alpha \urcorner \sqsupseteq \hat{F}(\alpha,\varepsilon^-)\label{eq8preuvesti}\end{equation}
Since $\varepsilon$ is an empty trace and $\hat{F}$ is well-formed in ${\cal{C}}$, then: \begin{equation}\hat{F}(\alpha,\varepsilon^-)=\ulcorner \alpha \urcorner\label{eq8preuvestiih}\end{equation}
From \ref{eq8preuvesti} and \ref{eq8preuvestiih} we have:
\begin{equation}\forall \alpha \in {\cal{A}}(\rho^+).F(\alpha,\rho^+) \sqcap \ulcorner \alpha \urcorner \sqsupseteq \ulcorner \alpha \urcorner\label{eq8preuvestiModMod1}\end{equation}
Since ${\cal{L}}^\sqsupseteq$ is a lattice, we have:
\begin{equation}\forall \alpha \in {\cal{A}}(\rho^+).F(\alpha,\rho^+) \sqsupseteq F(\alpha,\rho^+) \sqcap \ulcorner \alpha \urcorner\label{eq8preuvestiModMod2}\end{equation}
From \ref{eq8preuvestiModMod1} and \ref{eq8preuvestiModMod2}, we have: \begin{equation}\forall \alpha \in {\cal{A}}(\rho^+).F(\alpha,\rho^+) \sqsupseteq \ulcorner \alpha \urcorner\label{eq9preuvesti}\end{equation}
We deduce from \ref{eq6preuvesti} and \ref{eq9preuvesti}:
\begin{equation}\forall \alpha \in {\cal{A}}(m).(F(\alpha,m) \sqsupseteq \ulcorner \alpha \urcorner)  \vee (\ulcorner K(I) \urcorner \sqsupseteq \ulcorner \alpha \urcorner)\label{eq10preuvesti}\end{equation}
\end{itemize}
\end{itemize}
\begin{itemize}
\item [$\bullet$]We assume for $|\rho|\le n$, we have:
\begin{equation}\forall \alpha \in {\cal{A}}(m).(F(\alpha,m) \sqsupseteq \ulcorner \alpha \urcorner)  \vee (\ulcorner K(I) \urcorner \sqsupseteq \ulcorner \alpha \urcorner)\label{eq11preuvesti}\end{equation}
Let's prove that for $|\rho|=n+1$, we have:
\begin{equation}\forall \alpha \in {\cal{A}}(m).(F(\alpha,m) \sqsupseteq \ulcorner \alpha \urcorner)  \vee (\ulcorner K(I) \urcorner \sqsupseteq \ulcorner \alpha \urcorner)\label{eq12preuvesti}\end{equation}
Let $m \in {\cal{M}}$ such that $\rho^+ \models_C m $ with $|\rho|=n+1$.\\
\\
Let $\rho=\rho_n.e$\\
  \\
Since  $|\rho|=n+1$, we have: \begin{equation}|\rho_n|\leqslant n \mbox{ and } |e|\leqslant n \label{eq13preuvesti}\end{equation}
  From \ref{eq11preuvesti} and \ref{eq13preuvesti} and since $\rho_n^+\models_{\cal{C}}\rho_n^+$ and $e^+\models_{\cal{C}}e^+$ we have:
  \begin{equation} \forall \alpha \in {\cal{A}}(\rho_n^+).(F(\alpha,\rho_n^+) \sqsupseteq \ulcorner \alpha \urcorner)  \vee (\ulcorner K(I) \urcorner \sqsupseteq \ulcorner \alpha \urcorner) \label{eq14preuvesti}\end{equation}
  $$\mbox{and}$$ \begin{equation} \forall \alpha \in {\cal{A}}(e^+).(F(\alpha,e^+) \sqsupseteq \ulcorner \alpha \urcorner)  \vee (\ulcorner K(I) \urcorner \sqsupseteq \ulcorner \alpha \urcorner) \label{eq15preuvesti}\end{equation}
  From \ref{eq14preuvesti}, \ref{eq15preuvesti} and since $F$ is well-formed and ${\cal{L}}^\sqsupseteq$ is a lattice we have:
  \begin{equation} \forall \alpha \in {\cal{A}}(\rho_n^+\cup e^+).(F(\alpha,\rho_n^+)\sqcap F(\alpha,e^+) \sqsupseteq \ulcorner \alpha \urcorner)  \vee (\ulcorner K(I) \urcorner \sqsupseteq \ulcorner \alpha \urcorner) \label{eq16preuvesti}\end{equation}

Since $F$ is full-invariant-by-intruder in ${\cal{C}}$ and $\rho^+\models_{\cal{C}} m$, then: \begin{equation}\forall \alpha \in {\cal{A}}(m). (F(\alpha,m)\sqsupseteq F(\alpha,\rho^+))\vee (\ulcorner K(I) \urcorner \sqsupseteq \ulcorner \alpha \urcorner) \label{eq17preuvesti}\end{equation}
  Since: \begin{equation}\rho^+=\rho_n^+ \cup e^+\label{eq18preuvesti}\end{equation}
 Then from \ref{eq18preuvesti} and since ${\cal{L}}^\sqsupseteq$ is a lattice we have:
  \begin{equation}\forall \alpha \in {\cal{A}}(\rho^+). F(\alpha,\rho^+) = F(\alpha,\rho_n^+) \sqcap F(\alpha,e^+) \label{eq19preuvesti}\end{equation}
 From \ref{eq16preuvesti} and \ref{eq19preuvesti}  we have:
  \begin{equation}\forall \alpha \in {\cal{A}}(\rho^+). (F(\alpha,\rho^+)\sqsupseteq \ulcorner \alpha \urcorner)\vee (\ulcorner K(I) \urcorner \sqsupseteq \ulcorner \alpha \urcorner) \label{eq20preuvesti}\end{equation}
  From \ref{eq17preuvesti} and \ref{eq20preuvesti} we have:\\
  \begin{equation}\forall \alpha \in {\cal{A}}(m). (F(\alpha,m)\sqsupseteq \ulcorner \alpha \urcorner)\vee (\ulcorner K(I) \urcorner \sqsupseteq \ulcorner \alpha \urcorner) \label{eq21preuvesti}\end{equation}

\end{itemize}
  From \ref{eq10preuvesti}, the induction assumption in \ref{eq11preuvesti} and from \ref{eq21preuvesti}, we have: for all $m \in {\cal{M}}$ and all $\rho \in [\![p]\!]$:
  \[
  \rho \models_{\cal{C}} m \Rightarrow (\forall \alpha \in {\cal{A}}(m). (F(\alpha,m)\sqsupseteq \ulcorner \alpha \urcorner)\vee (\ulcorner K(I) \urcorner \sqsupseteq \ulcorner \alpha \urcorner))
  \]
  Or else:
  \[
\forall m \in {\cal{M}}. [\![p]\!] \models_{\cal{C}} m \Rightarrow \forall \alpha \in {\cal{A}}(m).(F(\alpha,m)\sqsupseteq \ulcorner \alpha \urcorner)\vee (\ulcorner K(I) \urcorner \sqsupseteq \ulcorner \alpha \urcorner)
\]

\end{preuve}

\begin{thm}(Correctness of increasing protocols)\label{mainThmouchArticleProofs}
$ $\\
Let $F$ be a ${\cal{C}}$-reliable function and $p$ a $F$-increasing protocol.
\begin{center}
$p$ is ${\cal{C}}$-correct with respect to the secrecy property.
\end{center}\index{Confidentialité}

\end{thm}
$ $\\
See the proof \ref{PR5}

\begin{preuve}\label{PR5} 
$ $\\
Let's assume that $p$ reveals an atomic secret $\alpha$. \\
\\
From the definition \ref{divulgationProofs} we have:
\begin{equation}\exists \rho\in [\![p]\!].(\rho \models_{\cal{C}} \alpha)\wedge(\ulcorner K(I) \urcorner \not \sqsupseteq \ulcorner \alpha \urcorner) \label{preuvthe1}\end{equation}
Since $F$ is a  ${\cal{C}}$-reliable function and $p$ an $F$-increasing protocol, we have from the lemma \ref{lemprincProofs}:
\begin{equation}(F(\alpha,\alpha)\sqsupseteq \ulcorner \alpha \urcorner)\vee (\ulcorner K(I) \urcorner \sqsupseteq \ulcorner \alpha \urcorner)\label{preuvthe2}\end{equation}
From \ref{preuvthe1} and \ref{preuvthe2}, we have:
\begin{equation}F(\alpha,\alpha)\sqsupseteq \ulcorner \alpha \urcorner \label{preuvthe3}\end{equation}
As $F$ is well-formed in ${\cal{C}}$, then:
\begin{equation}F(\alpha,\alpha)= \bot \label{preuvthe4}\end{equation}
From \ref{preuvthe3} and \ref{preuvthe4} we have:
\begin{equation}\bot= \ulcorner \alpha \urcorner \label{preuvthe5cor}\end{equation}
\ref{preuvthe5cor} is not possible because it is opposite to:  $\ulcorner K(I) \urcorner \not \sqsupseteq \ulcorner \alpha \urcorner$ in \ref{preuvthe1}.\\
\\
Then $p$ is ${\cal{C}}$-correct for the property of secrecy.
\end{preuve} 